\documentclass[twocolumn,epsf,psfig]{revtex4-2}
\usepackage{amssymb}
\usepackage{epsfig}
\DeclareMathAlphabet{\pazocal}{OMS}{zplm}{m}{n}            
\usepackage{graphicx}
\usepackage{color}
\usepackage{bm}
\usepackage[normalem]{ulem}     
\usepackage{soul}
\usepackage{amsmath}
\usepackage{braket} 
\usepackage{rotating}
\usepackage{multirow} 
\DeclareMathAlphabet{\pazocal}{OMS}{zplm}{m}{n}            
\usepackage{graphicx}
\usepackage{hyperref}

\begin{document}
\title{Magnetic octupoles as the order parameter for unconventional antiferromagnetism}     

\author{Sayantika Bhowal} 
\affiliation{Materials Theory, ETH Zurich, Wolfgang-Pauli-Strasse 27, 8093 Zurich, Switzerland} 

\author{Nicola A. Spaldin}
\affiliation{Materials Theory, ETH Zurich, Wolfgang-Pauli-Strasse 27, 8093 Zurich, Switzerland}

\date{\today}

\begin{abstract}
We show that time-reversal symmetry broken, centrosymmetric antiferromagnets with non-relativistic spin-splitting are conveniently
described in terms of the ferroic ordering of magnetic octupoles. The magnetic octupoles are the lowest-order ferroically ordered
magnetic quantity in this case, and so are the natural order parameter for the transition into the magnetically ordered state.
They provide a unified description of the broken time-reversal symmetry and the non-relativistic spin splitting as well as a
platform for manipulating the latter, and account for other phenomena, such as piezomagnetism, characteristic of this class of
antiferromagnets. Unusually for antiferromagnets, we show that the magnetic octupoles cause a non-zero magnetic Compton scattering,
providing a route for their direct experimental detection. We illustrate these concepts using density-functional and model
calculations for the prototypical non-relativistic spin-split antiferromagnet, rutile-structure manganese difluoride, MnF$_2$.

\end{abstract}

\maketitle

\section{Introduction}

The behavior that we now know as antiferromagnetism was first noticed around 100 years ago, when peaks in both specific heat and magnetic susceptibility were observed in materials such as MnO that have zero net magnetic moment \cite{Millar:1928,Tyler:1933}. Soon after, N\'eel proposed a model in which local magnetic dipole moments of equal magnitude on two sublattices order in an antiparallel fashion \cite{Neel:1936}. While the predictions of the N\'eel  model were consistent with the observations \cite{Bizette/Squire/Tsai:1938}, another ten years elapsed before neutron diffraction provided the first direct evidence of antiferromagnetic ordering of magnetic dipoles \cite{Shull/Smart:1949}. 

Usually the order parameter, $\vec{L}$, of an antiferromagnet (AFM) is defined in terms of the difference between the local magnetic dipole moments, $\vec{M}_1$ and $\vec{M}_2$, on the two sublattices, $\vec{L} = \vec{M}_1 - \vec{M}_2$. Such a definition is conceptually intuitive, but lacks the convenience provided by ferroic order parameters such as the magnetization, $\vec{M}$ in ferromagnets or the electric polarization, $\vec{P}$ in ferroelectrics. For example, the antiferromagnetic vector does not provide information about the conjugate field required to select for a particular antiferromagnetic domain, and fails to distinguish between antiferromagnets that do or do not break time-reversal symmetry. The magnetic dipoles, however, are just one of the terms in a multipole expansion of the energy of a general magnetization density in a magnetic field. They are generally the lowest-order local multipole on an atomic site, which makes them appealing for classifying magnetic order, but there is no fundamental reason why they should necessarily be the best choice. In particular, when the magnetic dipoles order antiferromagnetically, higher order multipoles that order ferroically might be more suitable.

Indeed, such a higher-order-multipole description is now established in the case of  AFMs that break both time-reversal  ($\mathcal{T}$)  and space-inversion ($\mathcal{I}$) symmetries, and which are classified by the ferroic ordering of their local magnetoelectric multipoles \cite{ClaudeSpaldin2007,VanAken-et-al:2007,Spaldin2013}. The magnetoelectric multipoles make up the next-order term, beyond the magnetic dipoles, in the multipole expansion of the magnetic interaction energy (see Eqn.~\ref{hint}), and so depend linearly on both position $r$ and magnetic moment $\mu$. All such $\mathcal{T}$ and $\mathcal{I}$-broken AFMs therefore exhibit a linear magnetoelectric response, in which an applied electric field induces a magnetization linear in the field strength and vice versa \cite{Spaldin/Fiebig/Mostovoy:2008}. Their conjugate field is the product of electric and magnetic fields, which is exploited in so-called magnetoelectric annealing to select for a particular antiferromagnetic domain in magnetoelectric devices \cite{Binek/Doudin:2005,Borisov_et_al:2005}. The ferroic ordering of magnetoelectric multipoles also plays a crucial role in antiferromagnetic spintronics \cite{Watanabe2018,Florian2020} and skyrmionics \cite{Gobel2019,BhowalSpaldin2022}, and can give rise to unconventional transport properties \cite{GaoXiao2018}. 

\begin{figure}[t]
\centering
\includegraphics[width=\columnwidth]{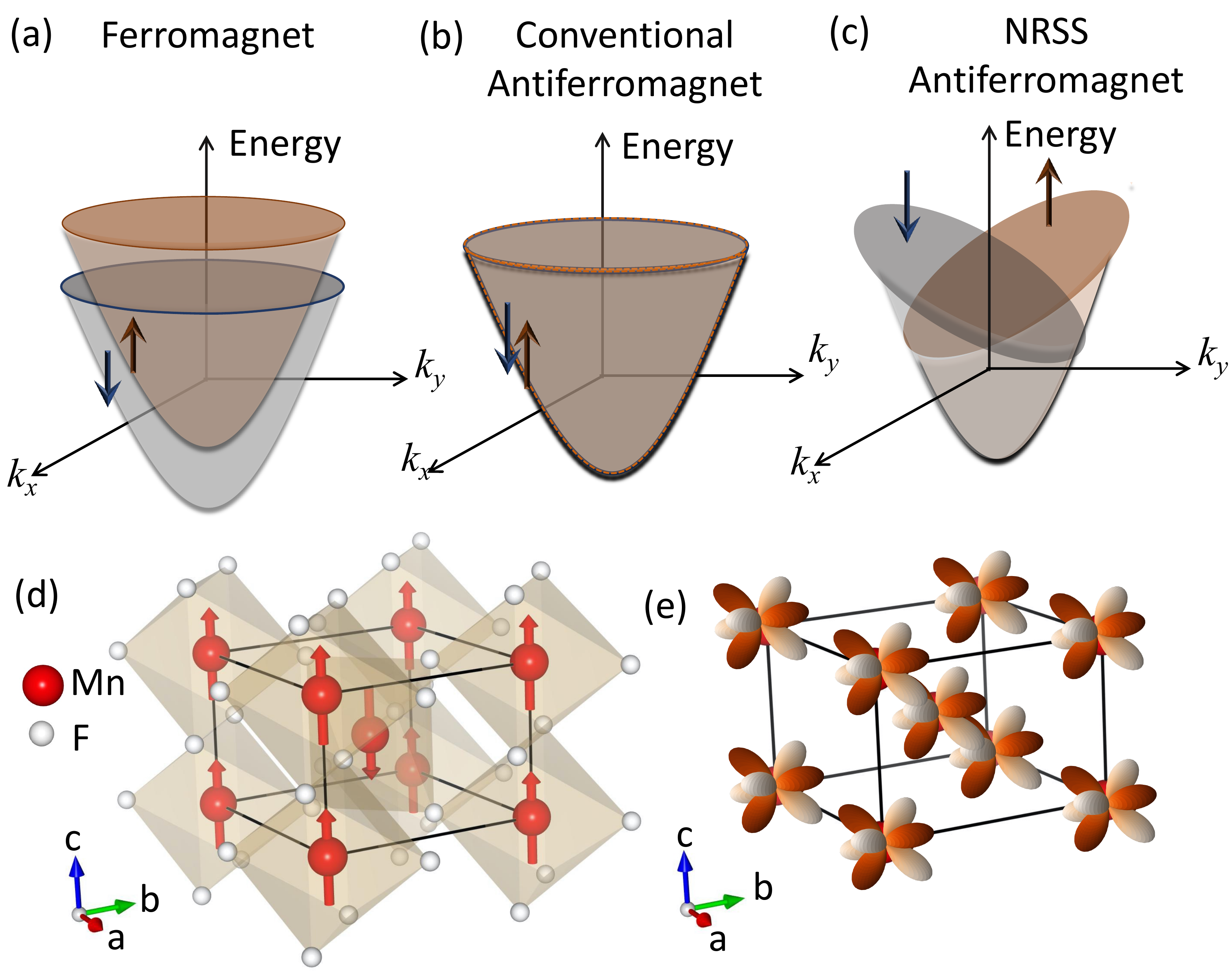}
 \caption{Non-relativistic spin-split (NRSS) antiferromagnets and ferro-type magnetic octupolar order. (a)-(c) shows the spin splitting of the bands for conventional ferromagnets (Zeeman splitting), antiferromagnets (degenerate spin-polarized bands), and the recently discovered AFMs with NRSS (symmetric in $\vec k$) of their bands. (d) and (e) show the antiferro-type magnetic dipolar (arrows, (d)) order  and ferro-type magnetic ${\cal O}_{32}^-$ octupolar (colored anisotropic octupolar magnetic distribution, (e)) order in MnF$_2$ respectively.    
 }
 \label{fig1}
 \end{figure}

Recently there has been renewed interest in a class of AFMs that break time-reversal symmetry and exhibit a spin-splitting of their energy bands that is not of relativistic origin (in conventional antiferromagnets the bands are doubly spin degenerate). First invoked in 1964 \cite{PekarRashba1964}, such non-relativistic spin-splitting (NRSS) is typically much larger than relativistic Rashba-like spin splitting, and can be substantial in materials containing only light elements.  An important recent development was the articulation of guiding principles for realizing such unconventional magnetism in materials \cite{Yuan2020,Yuan2021,Smejkal2022PRX,Yuan2022arXiv,YuanNov2022arXiv} so that proposed unconventional properties of both fundamental and technological importance \cite{Smejkal2022,Smejkal2022PRX}, including efficient spin-current generation \cite{Hernandez2021,Shao2021,Bose2022}, spin-splitting torque \cite{Bai2022,Karube2021}, giant magnetoresistance \cite{Libor2022}, spontaneous Hall effect \cite{Libor2020, Feng2020,SmejkalAHE2022} and superconductivity \cite{Mazin2022}, chiral magnons \cite{SmejkalNov2022arXiv} could now be within reach. A new name was even introduced -- altermagnet -- to describe this class of AFMs \cite{Smejkal2022} [see Fig. \ref{fig1} (a)-(c)]. Note that many NRSS AFMs are centrosymmetric and so are not described by ferroic ordering of magnetoelectric multipoles.

Here we show that such time-reversal symmetry broken, centrosymmetric AFMs are conveniently described in terms of the ferroic ordering of magnetic octupoles. The magnetic octupoles form the next term in the magnetic multipole expansion after the magnetoelectric multipoles, and are the lowest-order ferroically ordered magnetic quantity in this case. They are the natural order parameter for the transition into the magnetically ordered state, and provide a convenient and unified description of the broken $\mathcal{T}$ symmetry and the non-relativistic spin splitting. Importantly for potential device applications, they provide a platform for manipulating the spin-splitting. They also account for other phenomena displayed by this class of AFMs such as the piezomagnetic effect \cite{Baruchel1988,Ma2021} and strong magnetic anisotropy \cite{Mazin2021}, and allow us to predict new behaviors such as an anti-piezomagnetism. Finally, we show that, unusually for an antiferromagnet, they will have a non-zero magnetic Compton scattering, providing a route for their direct experimental detection. 

We illustrate our ideas using rutile-structure manganese difluoride, MnF$_2$. MnF$_2$ has been widely explored as a classic example of a two sub-lattice AFM  over the past century \cite{deHaas1940,Seehra1984,Yamani2010}, and was recently identified as a prototype centrosymmetric AFM with NRSS \cite{Yuan2020}. Importantly, Mn ions of opposite spin orientation have inequivalent fluorine environments (Fig. \ref{fig1} (d)). This results in identical, ferroically ordered ${\cal O}_{32^-}$ magnetic octupoles at each Mn site (Fig. \ref{fig1} (e)), which cause the broken $\mathcal{T}$ symmetry in spite of the AFM spin compensation.

The remaidner of the manuscript is organized as follows. We begin by briefly describing the crystal and magnetic structures of MnF$_2$ in Section \ref{structure}. This is followed by our discussion of the ordered magnetic octupoles and their role in NRSS in Section \ref{octupole}. In Section \ref{implication} we predict new behaviors resulting from the ferromagneto-octupolar order that await experimental verification, and propose magnetic Compton scattering as a route to the direct detection of magnetic octupoles. Finally, we summarize our results and discuss promising future directions in Section \ref{discussion}.

\section{Crystal and Magnetic Structure} \label{structure}

MnF$_2$ crystallizes in the centrosymmetric tetragonal rutile structure with the space group symmetry $P4_2/mnm$ ($D_{4h}$ point group) \cite{Baur1971,Yuan2020}. As depicted in Fig. \ref{fig1} (d), the unit cell contains two formula units, with two Mn atoms at the corner and the center of the unit cell, octahedrally coordinated by the F atoms. Importantly, the F environment around the Mn atom at the center is rotated by 90$^\circ$ around the $z$ axis with respect to that at the corner Mn atom. As a result of this non-equivalent F environment, the Mn sites, although equivalent, are not related by a lattice translation. This has a crucial impact on the symmetry of the AFM structure of MnF$_2$ (magnetic space group $P4_{2}~'/mnm'$) below the N{\'e}el temperature (T$_{\rm N} = 67$ K \cite{Stout1942}), where the collinear Mn spins align antiparallely along [001] \cite{Erickson1953} [see Fig. \ref{fig1} (d)]. Such a magnetic configuration  breaks the $\cal T$ symmetry despite the vanishing magnetization, since time-reversal plus translation is not a symmetry of the antiferromagnetic configuration. 

\section{Magnetic Octupole and Non-relativistic Spin Splitting } \label{octupole} 

The broken $\cal T$ symmetry in the absence of any net magnetization is indicative of the existence of magnetic multipole of higher order than the magnetic dipole. Such multipoles appear in the expansion of the interaction energy ${\cal E}_{\rm int}$ between a spatially varying magnetic field $\vec H(\vec r)$ and a magnetization density $\vec \mu(\vec r)$ \cite{Spaldin2008,Spaldin2013}, 
\begin{widetext}
\begin{eqnarray} \nonumber  \label{hint}
{\cal E}_{\rm int} &=& - \int  \vec \mu(\vec r) \cdot \vec H(\vec r) d^3r \\ 
                            &=& \underbrace{- \int  \vec \mu(\vec r) \cdot \vec H(0) d^3r}_{\text{\normalfont Dipolar~term}} \underbrace{-  \int  r_i \mu_j (\vec r) \partial_i H_j (0) d^3r}_{\rm Magnetoelectric~ multipolar~ term} \underbrace{- \int  r_i r_j\mu_k (\vec r) \partial_i \partial_j H_k (0) d^3r}_{\rm Octupolar~term}~...~~~.
\end{eqnarray}
\end{widetext}
In a compensated AFM, the net magnetization $ \vec {M}= \int  \vec \mu(\vec r) d^3r$ is absent and so the conventional dipolar Zeeman term which is the first term in the above expansion, does not contribute. Furthermore, the presence of inversion symmetry in centrosymmetric antiferromagnets with symmetric NRSS with respect to $\vec k$ forbids the existence of any ferro-type magnetoelectric multipole ${\cal M}_{ij} =  \int  r_i \mu_j (\vec r) d^3r$ since these break inversion symmetry, forming the second term in Eq. (\ref{hint}). This makes the first symmetry-allowed ferro-type magnetic multipole the inversion-symmetric rank-3 magnetic octupole, ${\cal O}_{ijk} =  \int  r_i r_j \mu_k (\vec r) d^3r$, which forms the third term in the above expansion. While this simple symmetry argument suggests that the magnetic octupole is the first allowed net nonzero magnetic multipole in a centrosymmetric AFM with NRSS, its existence can only be confirmed through an explicit computational  analysis of the multipoles. 

In the following we take  MnF$_2$ as a representative system for such unconventional AFMs and explicitly analyze the multipoles in the system. In particular, we focus on the magnetic octupole,  show its possible manipulation via structural and magnetic modifications, and correlate it to the characteristic non-relativistic spin splitting of the energy bands.

\subsection{Multipole Analysis} \label{multipole}

In order to compute the atomic-site multipoles, we decompose the density matrix $\rho_{lm,l'm'}$, computed using density-functional theory as implemented in the Elk code (see the computational details in Appendix A), into tensor moments \cite{Spaldin2013}. Since the desired parity even multipoles (as the structure has inversion symmetry) have contributions from even $l+l'$ terms, we evaluate both $d-d$ and $p-p$ matrix element contributions. We consider both $\cal T$ even (charge) and odd (magnetic) multipoles.

\begin{figure}[t]
\centering
\includegraphics[width=\columnwidth]{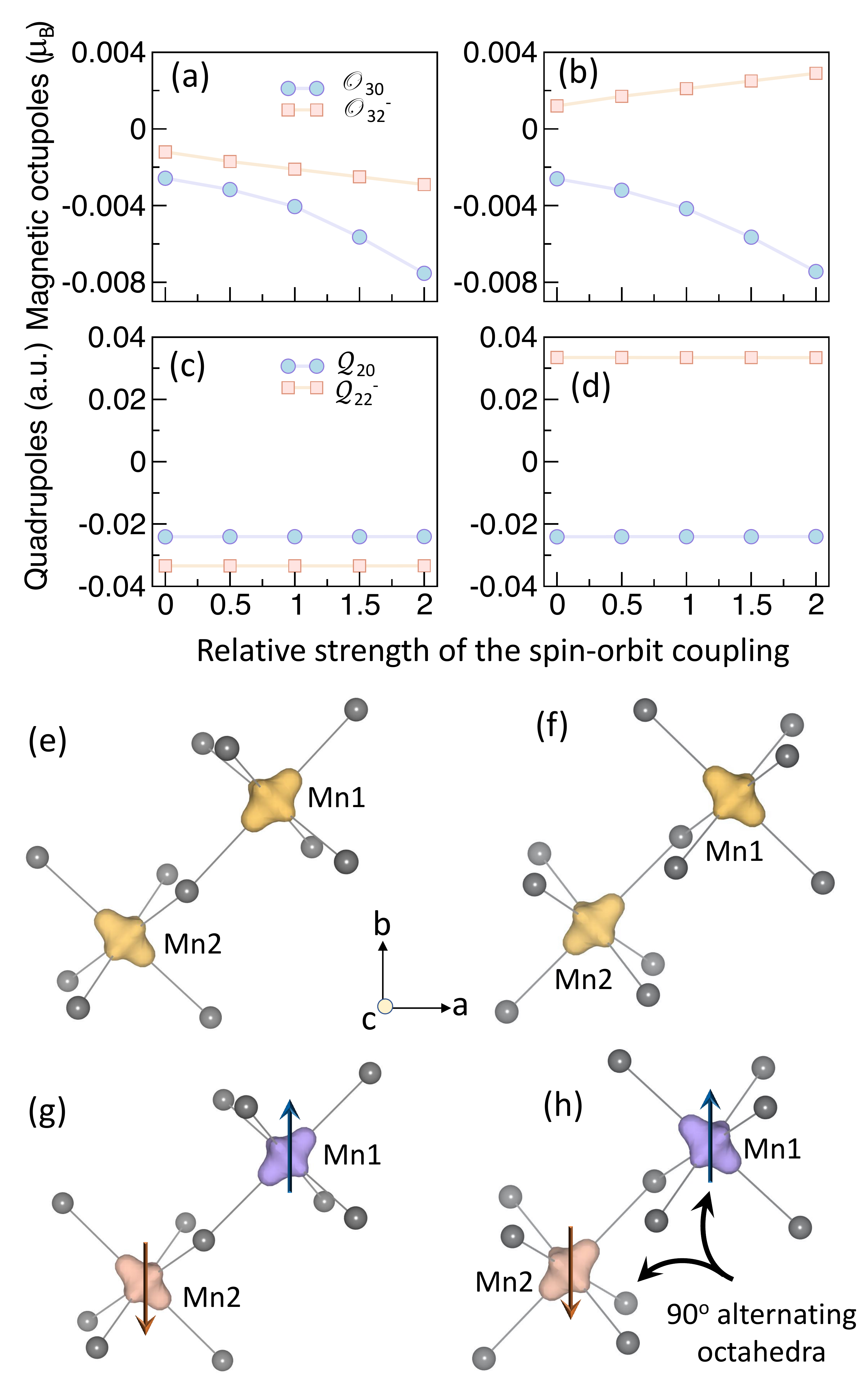}
 \caption{ Variation of magnetic octupoles at the Mn atoms as the relative strength of the spin-orbit coupling $\lambda_r$ is varied for the (a) crystal structure of MnF$_2$ and (b) for the modified structure. The same variation of the charge quadrupoles for the (c) crystal structure of MnF$_2$ and (d) for the modified structure. Here $\lambda_r=\lambda_f/\lambda$, with $\lambda_f$ and $\lambda$ being the enforced value of the spin-orbit coupling constant in the calculation and its actual value in the material respectively. The band-decomposed charge densities for the top valence band in Fig. \ref{fig3} (a) of MnF$_2$ in the $a$-$b$ plane for the (e) crystal structure of MnF$_2$ and (f) for the modified structure. (g) and (h) are the same, showing the band-decomposed magnetization densities. The opposite MnF$_6$ octahedral rotations are indicated in black arrows.
 }
 \label{fig2}
 \end{figure}

The computed magnetic octupoles, ${\cal O}_{32^-}$ and ${\cal O}_{30}$, and electric quadrupoles, ${\cal Q}_{22^-}$ and  ${\cal Q}_{20}$, are shown in Fig. \ref{fig2} (a) and (c) as the relative strength of the spin-orbit coupling constant, $\lambda_r$, is varied. As we can see from this variation, the magnetic octupoles are non-zero even without the presence of the spin-orbit coupling. It is also clear from Fig. \ref{fig2} that the magnitudes of the magnetic octupoles depend on $\lambda_r$, whereas the values of the quadrupoles remain invariant, suggesting that the quadrupoles have only structural origin while the magnetic octupoles may have both structural and magnetic dependencies. We note that a pure structural origin of quadrupoles is not a general case for any systems with quadrupolar distortion, e.g., the quadrupoles in the iso-space-group compound Ba$_2$MgReO$_6$ (with a canted antiferromagnetic spin configuration) are reported to have a strong spin dependence \cite{Mansouri2021}.  

The computed magnetic multipoles show the presence of a ferro-type magnetic octupole ${\cal O}_{32^-}$ (Fig. 1e) with real-space representation $xym_z$ (where $m_z$ is the $z$ component of the magnetic moment) at the Mn sites, belonging to the totally symmetric irreducible representation $A_{1g}$. The existence of the magnetic octupole is consistent with the $B_{1g}^-$ active representation of MnF$_2$ as well as the symmetry analysis described earlier. The magnetic octupole ${\cal O}_{30}$ with real-space representation $(3z^2-r^2)m_z$, also has a non-zero value at the Mn sites, however, they have anti-parallel alignment between the Mn sites, resulting in an absence of net ${\cal O}_{30}$ octupole moment. 
We also found non-zero magnetic octupole components ${\cal Q}^{(\tau)}_{x^2-y^2}$ and $t^{(\tau)}_{z}$ with ferro and antiferro-type alignments respectively, which we will discuss later in Section \ref{PM} and in Appendix D. In addition to these magnetic octupoles, the crystal structure of MnF$_2$ also hosts electric quadrupoles, ${\cal Q}_{20}$ with $(3z^2-r^2)$ distortion and ${\cal Q}_{22^-}$, representing the $xy$-structural distortion, which have ferro- and antiferro-type alignments respectively. 

For a physical understanding and better visualization, we further compute the band-decomposed charge and spin densities for the top valence band (which also undergoes NRSS) in the electronic structure of MnF$_2$, shown in Fig. \ref{fig3} (a) and the results are shown in Fig. \ref{fig2} (e) and (g). As apparent from the figure, the charge density around the Mn atoms is highly anisotropic in the $x-y$ plane, a signature of the existing ${\cal Q}_{22^-}$ quadrupole. Interestingly, the spin density around the Mn atoms, shown in Fig. \ref{fig2} (d), follows the anisotropic charge density, indicating a correlation between the spin anisotropy (quantified by the magnetic octupoles) and the charge anisotropy (quantified by the electric quadrupoles). This further justifies the dependence of the octupoles on $\lambda_r$. 

\subsection{Magnetic Octupolar Domains: Correlation to Structure and Spin} \label{manipulation}

We now show how the coupling between magnetic octupoles and charge quadrupoles determine the magnetic octupolar domains.  Since the magnetic octupoles are linked to the NRSS, as we show in the next section, the  understanding of the magnetic octupolar domain is also useful in manipulating the NRSS.

We begin by changing the fluorine environment around the Mn atoms, without affecting the spin arrangements at the Mn sites. Specifically, we change the coordinates of the F ions from $ 4f: (x,x,0) \rightarrow 4g: (x, -x,0)$, which alters the F-Wyckoff site symmetry from $4f$ to $4g$, while keeping the space-group symmetry unchanged. We note that structurally (without considering the magnetism), the new structure is equivalent to the original crystal structure of MnF$_2$ (shown in Fig. \ref{fig1} (c)), with a shifted origin at (0.5,0.5,0.5), so that the central and corner Mn atoms are exchanged in the new structure. This results in a 90$^\circ$ alternation of MnF$_6$ octahedral rotation and, hence, their distortion in the $x-y$ plane (see Figs. \ref{fig2} (e)-(h)). Correspondingly, this leads to a reversal of sign in the computed antiferro-type ${\cal Q}_{22^-}$ quadrupole for the modified structure, as depicted in Fig. \ref{fig2} (d). This is also evident from the changes in the anisotropic charge density distribution around the Mn atoms in the modified structure as shown in Fig. \ref{fig2} (f). Note that the modified fluorine environment, however, does not affect the distortion of the MnF$_6$ octahedra along $z$ direction, and the sign of the ferro type ${\cal Q}_{20}$ quadrupole, therefore, remains unaltered (see Fig. \ref{fig2} (d)).    

In order to see the impact of the charge quadrupoles on the magnetic octupoles, we further analyze the magnetic multipoles of this modified structure. Interestingly, our computed octupoles show that the sign of the ${\cal O}_{32^-}$ octupole reverses whereas that of ${\cal O}_{30}$ remains as it is (see Fig. \ref{fig2} (b)), showing the reversal of the magnetic octupolar domain. Corresponding changes in the anisotropic magnetization density around the Mn atoms are shown in Fig. \ref{fig2} (h). This also emphasizes the correlation between ${\cal O}_{32^-}$ octupole and ${\cal Q}_{22^-}$ quadrupole; and ${\cal O}_{30}$ octupole and ${\cal Q}_{20}$ quadrupole. Note that the ${\cal O}_{32^-}$ and ${\cal O}_{30}$ octupoles remain  ferro- and antiferro-type respectively. This suggests selecting the magnetic octupolar domain by only changing the position of the non-magnetic F atoms, without affecting the magnetic Mn atom's position or its spin arrangements, emphasizing a strong interplay between lattice and the magnetic configuration, quantified by the magnetic octupole. 

We close this section by describing the manipulation of the magnetic octupole by flipping the direction of all the Mn-spins while keeping the same antiferromagnetic arrangement between the Mn atoms. Physically, it corresponds to a different antiferromagnetic domain. The reversal of the Mn-spins results in a reversal of the sign of both octupoles, in contrast to the previous case where only the ferro-type ${\cal O}_{32^-}$ octupole reverses its sign. Note that, in this case electric quadrupoles remain the same, as they do not depend on the spin arrangement. It is interesting to point out that this manipulation of the magnetic octupole has in fact important physical implications. For example, the two antiferromagnetic domains with ferro-type octupoles of opposite sign can also be visualized as two separate ferro-octupolar domains. Such re-visualization is particularly useful in describing important physical properties that are characteristics of the magnetic octupoles as well as understanding the conjugate fields for creating such ferro-octupolar domains, as we discuss later in Section. \ref{implication}.  

\subsection{Relevance to Non-relativistic Spin-splitting}

Next we link the unconventional spin-splitting of the energy bands in the Brillouin zone (BZ) of the antiferromagnetic MnF$_2$ to the ferro-octupolar order using the reciprocal-space representation of the ferro-type octupole. Since the magnetic octupole can be manipulated by modifying the F-environment or the spin arrangements, as discussed above, we show that these changes can also be used to manipulate the spin-splitting of the bands.  

\begin{figure}[t]
\centering
\includegraphics[width=\columnwidth]{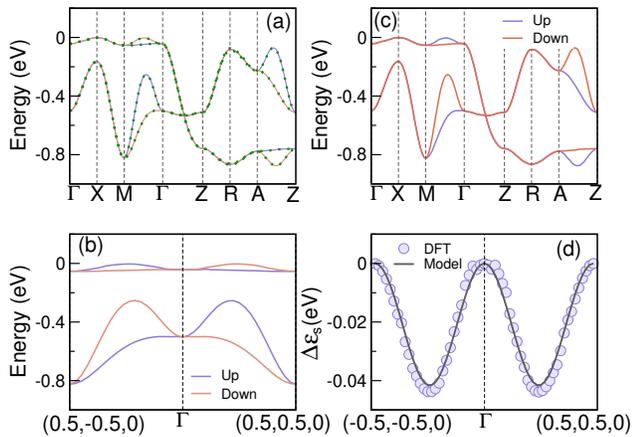}
 \caption{Spin splitting in MnF$_2$. (a) Band structure of MnF$_2$ in both absence and presence of spin-orbit interaction, depicting the spin splitting along $M \rightarrow \Gamma$ direction. The up and down spin-polarized bands in absence of spin-orbit coupling are shown in solid blue and red lines respectively and the bands in presence of spin-orbit interaction are indicated in green dots. (b) The corresponding band structure along the same high-symmetry $k$ path for the hypothetical modified structure (see text for details), showing the reversal of spin splitting along $M \rightarrow \Gamma$. (c) Band structure of MnF$_2$ showing the reversal of spin splitting as the momentum direction changes from [1$\bar{1}$0] to [110]. (d) A comparison of the DFT result and the tight-binding analytical expression, Eq. (\ref{energy}), for the energy splitting between the two top-most spin-polarized bands in (a). 
 }
 \label{fig3}
 \end{figure}

Our calculated antiferromagnetic band structure of MnF$_2$ both in the presence and absence of spin-orbit interaction is depicted in Fig. \ref{fig3}(a). As we see from the band structure, there is a significant energy splitting between the up- and down-spin bands along $\Gamma \rightarrow$ M direction in the BZ of MnF$_2$. Interestingly, the splitting is present even without the spin-orbit interaction, and inclusion of spin-orbit interaction does not affect the energy splitting along that direction, in agreement with the reported band structure in Ref. \onlinecite{Yuan2020}. The splitting is large compared to the typical relativistic Rashba-type spin-splitting and does not require any broken inversion symmetry of the structure \cite{Yamauchi2019}. 

To understand this unconventional spin-splitting in MnF$_2$, we analyze the reciprocal-space representation of the ferro-type ${\cal O}_{32^-}$ octupole. 
The reciprocal space representations of the multipoles have been used successfully to describe the band asymmetries in the BZ of noncentrosymmetric materials \cite{Watanabe2018,BhowalSpaldin2021,Bhowal2022}.  In contrast to the odd-parity multipoles, for which the real and reciprocal-space representations are rather counter-intuitive, for the even-parity multipoles, such as the ${\cal O}_{32^-}$ magnetic octupole, that are relevant here the analysis is much more straightforward. The reciprocal space representation in this case can simply be obtained by replacing $\vec r \rightarrow \vec k$ so that the reciprocal-space representation of ${\cal O}_{32^-}$ octupole ($xym_z$ in real space) is $k_xk_ym_z$. This immediately explains the splitting between up and down spin-polarized bands, with the spin polarization along $\hat z$, along $[110]$ direction in the momentum space, e.g., $\Gamma \rightarrow$ M and A$\rightarrow$ Z directions in momentum space. Note that such splitting will occur along any momentum direction with non-zero $k_x$ and $k_y$. 

Interestingly, since the reciprocal space representation is an even function of $\vec k$, the resulting spin-splitting should also be symmetric, in contrast to the anti-symmetric spin splitting in the Rashba interaction. This indeed is the case e.g., in MnF$_2$ with identical spin splitting along $[110]$ and $[\bar{1}\bar{1}0]$ directions in the momentum space. In addition from the representation $k_xk_ym_z$, we also expect the spin-splitting to reverse as the direction in the momentum space changes from [110] to [1$\bar{1}$0] ($d$-wave spin splitting). Indeed, the computed DFT band structure depicts such reversal of spin-splitting under ${\cal C}_4$ rotation of the momentum direction, as shown in Fig. \ref{fig3} (b). The representation analysis, therefore, confirms that the ferro-type ordering of the ${\cal O}_{32^-}$ octupoles is responsible for the spin splitting of the energy bands, analogous to the spin splitting of bands in a conventional ferromagnet with ferro-type magnetic dipole. Note that the atomic-site magnetic octupole, discussed here, is distinct from the cluster and bond multipoles predicted by Hayami {\it et. al.} \cite{Hayami2019,HayamiPRB2020} for spin splitting in collinear antiferromagnets. The magnetic octupole description has the advantage that it naturally occurs in a magnetic multipole expansion and also describes the order parameter for such unconventional antiferromagnetism, as discussed above.

To further verify the role of ${\cal O}_{32^-}$ octupole in generating the spin-splitting, we analyze the spin-splitting of the bands for the case of structural modification, discussed in the previous section, for which the ${\cal O}_{32^-}$ octupole switches sign. As expected, in this case, the spin splitting also reverses (see Fig. \ref{fig3} (c)). Similar reversal of the spin-splitting also occurs for the opposite magnetic dipolar domain (not shown here), in which the magnetic ${\cal O}_{32^-}$ octupoles also switch sign.

\subsection{Role of microscopic parameters in the spin splitting}

Having shown that the ferro-type ordering of the magnetic octupoles generates the spin-splitting in MnF$_2$, to determine the role of different microscopic parameters, such as electronic hopping, exchange splitting, etc., on the strength of the spin splitting, we next carry out a low-energy tight-binding (TB) analysis. For this purpose, we construct a minimal four-band TB model in the Bloch function basis with the order of the basis
set in the sequence $\{ {\rm Mn1}-d_{xz}, {\rm Mn1}-d_{yz}, {\rm Mn2}-d_{xz}, {\rm Mn2}-d_{yz} \}$. The Hamiltonian reads as follows
\begin{equation} \label{TB}
    {\cal H}_{t} =  \alpha(\vec k) \mathbb{I}  + \beta(\vec k) \Sigma_z \otimes \sigma_x + \gamma(\vec k) \Sigma_x \otimes \sigma_0 +\delta(\vec k) \Sigma_x \otimes \sigma_x.
\end{equation}
Here, $\mathbb{I}$ is a $4\times4$ identity matrix, $\vec\Sigma$ and $\vec \sigma$ are the Pauli matrices in the sublattice bases of Mn1 and Mn2 and in the orbital bases of $d_{xz}$ and $d_{yz}$ respectively, and $\sigma_0$ is an identity matrix in the orbital bases. The choice of the orbitals is governed by the predominant orbital characters of the top pair of valence bands along $\Gamma \rightarrow$ M in the BZ of MnF$_2$ (see Fig. \ref{fig6} in Appendix C). The functions $\alpha(\vec k), \beta(\vec k), \gamma(\vec k)$, and $\delta(\vec k)$ are determined by the effective $d$-$d$ hopping parameters ($t_i$, $i=1,4$) and the on-site energies ($\varepsilon_i$, $i=1,2$) of the orbitals and their explicit functional forms are given below, 
\begin{eqnarray}\label{functions}\nonumber
\alpha(\vec k) &=& \varepsilon_1 +2t_1 \cos(k_zc) \\ \nonumber
\beta(\vec k)  &=& \varepsilon_2 +2t_2 \cos(k_zc) \\ \nonumber
\gamma(\vec k) &=& 8t_3\cos \Big(\frac{k_xa}{2} \Big)\cos\Big(\frac{k_ya}{2} \Big)\cos\Big(\frac{k_zc}{2} \Big) \\
\delta(\vec k) &=& -8t_4 \sin\Big(\frac{k_xa}{2} \Big)\sin\Big(\frac{k_ya}{2} \Big)\cos\Big(\frac{k_zc}{2} \Big).
\end{eqnarray}
Here, $a$ and $c$ are the lattice constants of the tetragonal structure. For simplicity, we consider electronic hoppings only up to second nearest neighbor and the realistic TB parameters are extracted from the DFT band structure of MnF$_2$ using the NMTO downfolding technique \cite{AndersenSaha-Dasgupta}.  

The diagonalization of the Hamiltonian in Eq. \ref{TB}, gives us the four energy eigenvalues, 
\begin{eqnarray}\label{ev} \nonumber
{\cal E}^-_{\pm} (\vec k) &=& \alpha (\vec k) - \{ \beta (\vec k)^2 +(\delta(\vec k) \pm \gamma(\vec k))^2 \}^{1/2} \\ 
{\cal E}^+_{\pm} (\vec k) &=& \alpha (\vec k) + \{ \beta (\vec k)^2 +(\delta(\vec k) \pm \gamma(\vec k))^2 \}^{1/2}.
\end{eqnarray}
Analysis of the corresponding eigenvectors shows that there is an energy splitting $\Delta {\cal E} = {\cal E}^-_+ (\vec k) - {\cal E}^-_- (\vec k)$ between bands of dominant Mn1 and Mn2 sublattice contributions. Note that such energy splitting between bands of different sublattice characters is present prior to including the effect of antiferromagnet exchange splitting $J$. We now show that the inclusion of $J$ translates the sublattice splitting of the bands into the spin splitting of the bands.  

To include the effect of the antiferromagnetic exchange, we rewrite the Hamiltonian (\ref{TB}) in the basis of  $\{ {\rm Mn1}-d_{xz} \uparrow, {\rm Mn1}-d_{yz} \uparrow, {\rm Mn2}-d_{xz} \uparrow, {\rm Mn2}-d_{yz} \uparrow, {\rm Mn1}-d_{xz} \downarrow, {\rm Mn1}-d_{yz} \downarrow, {\rm Mn2}-d_{xz} \downarrow, {\rm Mn2}-d_{yz} \downarrow \}$ and add the exchange term ${\cal H}^{\rm AFM}_{\rm ex} = J {\cal S}_z \otimes (\Sigma_z \otimes \sigma_0)$ to it. The full Hamiltonian is given by,
\begin{equation} \label{H}
 {\cal H} = {\cal S}_0 \otimes {\cal H}_t + J {\cal S}_z \otimes (\Sigma_z \otimes \sigma_0) ~~~~.
\end{equation}
Here, $\vec {\cal S}$ and ${\cal S}_0$ are the Pauli matrices and the identity matrix in the spin basis. The exchange splitting energy between up and down spin polarized bands, $2J \approx 5$ eV, extracted from the computed spin-polarized densities of states of MnF$_2$. 

By diagonalizing the Hamiltonian (\ref{H}), we obtain the energy eigenvalues and focus on the spin-polarized top most valence bands, with energies ${\cal E}_{\uparrow}$ and ${\cal E}_{\downarrow}$. We note that their eigenvalues are identical to those of ${\cal E}^-_\pm$ in Eq. (\ref{ev}), except that $\beta(\vec k) \rightarrow J+\beta(\vec k)$. Physically, this means that the two Mn sublattices, that primarily contribute to those bands, have opposite spin polarization in the presence of antiferromagnetism and, therefore, they lead to the spin splitting of the bands. The explicit analytical form of the energy splitting is given by,
\begin{eqnarray} \nonumber \label{energy}
 \Delta {\cal E}_{s} &=& {\cal E}_{\uparrow}-{\cal E}_{\downarrow} \\ \nonumber
 &=& \{ (J+ \beta)^2 +(\delta - \gamma)^2 \}^{1/2} - \{ (\beta +J) + (\delta + \gamma)^2 \}^{1/2} \\ 
 &\approx& \frac{32}{{\cal \epsilon}}t_3t_4 \sin(k_xa) \sin(k_ya).
\end{eqnarray}
Here, in obtaining the last equality we have used the fact that $\epsilon \gg (\delta +\gamma)$, where $\epsilon = J + \beta \approx J + \varepsilon_2+ 2t_2$, using Eq. \ref{functions} and ignoring terms that are second order in $k_z$ or higher. Note that this approximation in $\epsilon$ becomes exact in the $k_z = 0$ plane, which contains the desired $\Gamma \rightarrow$ M momentum direction of spin-splitting. For a realistic set of parameters (listed in Appendix B), we compare the analytical result in Eq. (\ref{energy}) to the DFT computed energy splitting of the spin-polarized bands. As depicted in Fig. \ref{fig3} (d), they agree reasonably with each other, suggesting that our minimal model captures the essential physics of the spin-splitting in MnF$_2$. 

We pause here and analyze the obtained analytical relation in Eq. (\ref{energy}) for the spin-splitting energy. First of all, it is clear from Eq. (\ref{energy}), that $\Delta {\cal E}_{s} (\vec k) = \Delta {\cal E}_{s} (-\vec k)$, i.e., it is symmetric in $\vec k$, but changes sign under $(k_x, k_y) \rightarrow (k_x, -k_y)$, consistent with the computed DFT bands. Secondly, we see that the splitting energy $\Delta {\cal E}_{s}$ depends directly on the inter-sublattice hopping parameters, $t_3$ (intra-orbital) and $t_4$ (inter-orbital) in the absence of which $\Delta {\cal E}_{s}$ vanishes. This emphasizes the crucial role of interaction between the two sublattices, which in combination with the antiferromagnetic exchange, generates the spin splitting. Physically, this indicates that a structure-spin correlation, a reminiscent of the magnetic octupole as discussed before, is responsible for the spin splitting. It is interesting to point out that the inter-sublattice hopping $t_4$ (as well as the product $t_3t_4$) in MnF$_2$ is a symmetric hopping and it changes sign as the direction of hopping changes from [11z] to [1$\bar 1$z] with $z \ne 0$ leading to symmetric spin splitting. This is analogous to the antisymmetric hopping in a nonmagnetic broken-inversion symmetric system that gives rise to Rashba-like antisymmetric spin splitting of the energy bands \cite{Bhowal2022}. Finally, the TB analysis also provides a microscopic understanding of the reversal of spin splitting described before. For the modified structure, the onsite energy $\varepsilon_2$ and the hopping $t_2$ change sign, leading to a sign change in $\beta$. While, the sign change does not affect the energy eigenvalues ${\cal E}^-_{\pm} (\vec k)$ in Eq. \ref{ev} (since the dependence on $\beta$ comes in even power), it reverses the dominant sublattice contributions in the corresponding eigenvectors (as also evident from the full DFT band structure, depicted in the appendix C), resulting in a reversal of sublattice splitting of bands. Since, the sublattice splitting later transforms into the spin splitting, this consequently leads to the reversal of the spin splitting. The reversal of spin-splitting for the other antiferromagnetic domain follows directly from the resulting sign change in the antiferromagnetic exchange $J$, which, in turn, alters the spin polarization of the bands.     

Overall, the TB analysis provides a crucial insight into the roles of different microscopic parameters in generating the unconventional spin splitting of the energy bands in MnF$_2$. The TB analysis further serves as a link between the proposed  ``modern" ferro-octupolar order and the conventional antiferromagnetic dipolar order.

\section{Implications of Magnetic Octupole}\label{implication}

The next step is to identify the implications of the existing magnetic octupole in determining the physical properties of a centrosymmetric AFM with NRSS as well as its possible direct measurements. Here, we point out (A) the resulting physical properties, piezo and anti-piezomagnetic effects and (B) the possible detection of magnetic octupoles using the magnetic Compton scattering effect. Once again, we take MnF$_2$ as our example material for illustration. We show that the existing ferro-type magnetic octupole ${\cal O}_{32^-}$ describes the well-known piezomagnetic effect in MnF$_2$. More interestingly, however, the knowledge of  the antiferro-type magnetic octupole ${\cal O}_{30}$ helps us to predict a previously unknown anti-piezomagnetic effect. In addition to the underlying fundamental physics and technological applications of these effects, we also propose magnetic Compton scattering as an experimental technique for the detection of the apparently {\it hidden} magnetic octupoles. The corresponding measurement set-up as a guidance for future experiments are also discussed.   

\subsection{Piezo and Anti-piezomagnetic Effects} \label{PM}

The piezomagnetic effect, describes changes in magnetization due to an applied stress or changes in shape due to an applied magnetic field. It is particularly promising for applications because it provides a means for manipulating magnetism via strain engineering in antiferromagnets. In addition, since it is a linear coupling in contrast to the quadratic coupling in the commonly used magnetostriction, it also allows for magnetization switching. The recent demonstration that the dynamically excited optical phonons can induce the symmetry-breaking lattice distortions required in the piezomagnetic effect\cite{Disa2020,Formisano2022,Formisano2022JPCM}, has revived interest. Such optically induced strain would overcome the limitation of a large mechanical strain and lead to practical applications in memory and spintronic devices. 

The piezomagnetic effect has been predicted and experimentally shown for some AFMs with NRSS \cite{Baruchel1988,Disa2020,Ma2021}. 
In this section, we show that the piezomagnetic effect is the result of ferroic ordering of magnetic octupoles, and illustrate our ideas for the specific example of MnF$_2$. In addition, we predict an antipiezomagnetic effect in MnF$_2$ resulting from the antiferro-type ${\cal O}_{30}$ magnetic octupole.

{\it General symmetry description-}
We begin by correlating the symmetries of the magnetic octupole and the piezomagnetic response. We note that both are rank-3 tensors and have the same symmetry, breaking time-reversal symmetry while keeping inversion symmetry intact. To correlate the elements of the piezomagnetic response to the magnetic octupole, we analyze the non-zero elements in the magnetic octupole tensor ${\cal O}_{ijk} = \int \mu_i r_j r_k d^3r$ following the tensor decomposition reported in Ref. \cite{Urru2022}. Note that in general $i,j,k$ are the dummy indices and to be consistent with the indices of the piezomagnetic response tensor $\Lambda_{ijk}$, here we associate the index $i$ to the magnetization and $j$ and $k$ to spatial coordinates so that the octupole ${\cal O}_{ijk}$ is symmetric under the exchange of $j$ and $k$ indices by construction.

The octupole ${\cal O}_{ijk}$ can be decomposed into a totally symmetric tensor ${\cal S}_{ijk}$ of dimension 10 and an 8 dimensional non-symmetric residue tensor ${\cal R}_{ijk}$, that account for the 18 independent elements of ${\cal O}_{ijk}$ \cite{Urru2022}. The totally symmetric tensor ${\cal S}_{ijk}$ can be further decomposed into a traceless totally symmetric part $\tilde{\cal S}_{ijk}$ and a trace part ${\cal T}_{ijk}$ of dimensions 7 and 3 respectively and the residue tensor ${\cal R}_{ijk}$ into two irreducible components $\tilde{\cal R}^{(5)}_{ijk}$ and $\tilde{\cal R}^{(3)}_{ijk}$ of dimensions 5 and 3 respectively, so that ${\cal O}_{ijk}=\tilde{\cal S}_{ijk}+{\cal T}_{ijk}+\tilde{\cal R}^{(5)}_{ijk}+\tilde{\cal R}^{(3)}_{ijk}$. The explicit forms of each of these irreducible components are given in Ref. \cite{Urru2022}. Note that the 7 independent components of the totally symmetric traceless part $\tilde{\cal S}_{ijk}$ can be built from the spherical harmonics with $l = 3$ and, hence, these components are often exclusively referred to as the magnetic octupole, in contrast to the entire ${\cal O}_{ijk}$ tensor.

We now explicitly consider the case of MnF$_2$, which is known to exhibit a piezomagnetic effect \cite{Dzialoshinskii1958,Borovikromanov1960,Baruchel1980,Baruchel1988} with the non-zero elements of the piezomagnetic response tensor $\Lambda_{ijk}$, $\Lambda_{xyz}=\Lambda_{yxz}\ne\Lambda_{zxy}$ so that, 
\begin{eqnarray} \label{response}
& & {\cal M}_x = \Lambda_{xyz} \sigma_{yz}, {\cal M}_y=\Lambda_{yxz} \sigma_{xz}, {\cal M}_z=\Lambda_{zxy} \sigma_{xy}, 
\end{eqnarray}
where $\vec {\cal M}$ is the magnetization that results from the application of shear stress $\sigma_{ij}$.
We show next that the non-zero components of the piezomagnetic response of MnF$_2$ correlate to the ferro-type ${\cal O}_{32^-}$ octupole.

Analyzing the different elements of $\tilde{\cal S}_{ijk}$, we see that the ${\cal O}_{32^-}$ octupole, which has a ferro-type ordering in MnF$_2$, appears only when $i=x, j=y, k=z$ and for the permutation of the indices. $\tilde{\cal S}_{ijk}$ being symmetric, all these elements are equal in magnitude. Similarly, analyzing the elements of other irreducible components, we find that the only other multipole that has a ferro-type ordering in MnF$_2$ is ${\cal Q}^{(\tau)}_{x^2-y^2}$, identified as the $x^2-y^2$ quadrupole component of the toroidal moment density $\tau (\vec r) = \vec r \times \vec \mu (\vec r)$. This leads to non-zero elements in the residue tensor $\tilde{\cal R}^{(5)}_{ijk}$, with $\tilde{\cal R}^{(5)}_{xyz}=\tilde{\cal R}^{(5)}_{yzx}\ne\tilde{\cal R}^{(5)}_{zxy}$. Combining the ferro-type magnetic octupole components in MnF$_2$, and the tensor decomposition of the the magnetic octupole ${\cal O}_{ijk}$, we obtain ${\cal O}_{xyz}={\cal O}_{yxz}\ne{\cal O}_{zxy}$. This nicely correlates with the symmetry allowed as well as experimentally observed components of piezomagnetic response for MnF$_2$ in Eq. \ref{response}, confirming the one-to-one correlation between the piezomagnetic response and the magnetic octupole tensor. 

Finally we also predict an {\it anti}-piezomagnetic response in MnF$_2$ due to the antiferro-type ${\cal O}_{30}$ octupole. 
Upon application of stress we expect an additional change in the Mn spin moments which is, however, opposite for the two Mn atoms so that their contributions cancel each other, leading to a zero net magnetization. Here, the tensor decomposition guides us in predicting which spin components will change due to an applied strain with a certain orientation. Therefore, we follow the same procedure as before and analyze first the elements of the symmetric traceless $\tilde{\cal S}_{ijk}$ to identify the elements of $\tilde{\cal S}_{ijk}$ in which the ${\cal O}_{30}$ octupole appears.  These are $\tilde{\cal S}_{xxz} = \tilde{\cal S}_{yyz} \ne \tilde{\cal S}_{zzz}$. The elements with symmetric permutation of these indices are also allowed. For these same elements of the residue tensors $\tilde{\cal R}^{(5)}_{ijk}$ and $\tilde{\cal R}^{(3)}_{ijk}$, we find that only $\tilde{\cal R}^{(3)}_{xxz}=\tilde{\cal R}^{(3)}_{yyz} \ne \tilde{\cal R}^{(3)}_{zxx}=\tilde{\cal R}^{(3)}_{zyy}$ are non-zero due to the existence of the  antiferro-type multipole $t_z^{(\tau)}$, 
defined as the $z$ component of the moment of the toroidal moment density, in MnF$_2$. This means that if both ${\cal O}_{30}$ and $t_z^{(\tau)}$ were ferro-type, we would have the following non-zero components in the piezomagnetic response
\begin{eqnarray}
\Lambda_{xxz}=\Lambda_{yyz}, \Lambda_{zxx}=\Lambda_{zyy}, \text{and}~~ \Lambda_{zzz}~~.
\end{eqnarray}
However, since in fact ${\cal O}_{30}$ and $t_z^{(\tau)}$ have antiferro-type arrangement in MnF$_2$, the first equality in the above equation instead indicates that a spin component along $\hat x$ ($\hat y$) will develop at individual Mn sites if we apply a shearing stress $\sigma_{xz}$ ($\sigma_{yz}$) to the structure, with the developed spin components having an anti-parallel alignment between the Mn sites, so that there is no net magnetization along $\hat x$ ($\hat y$). We refer to this effect as an {\it anti}-piezomagnetic effect due to the generation of anti-parallel spin components upon application of stress in analogy to the piezomagnetic effect where parallel spin moments are generated to give rise to a net change in magnetization. 

\begin{figure}[t]
\centering
\includegraphics[width=\columnwidth]{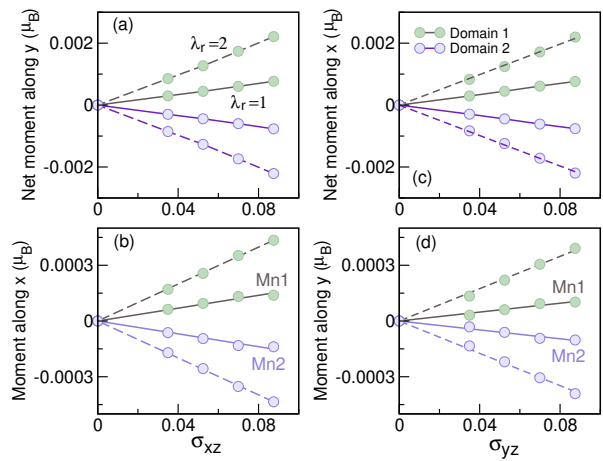}
 \caption{ Piezo- and antipiezo-magnetic effects in MnF$_2$. The variation of (a) the net magnetic moment along the $y$ direction and (b) the individual Mn magnetic moments along $x$ direction as the shear strain $\sigma_{xz}$ is varied. The variations of (c) the net magnetic moment along $x$ direction and (d) the individual Mn magnetic moments along $y$ direction as a function of the shear strain $\sigma_{yz}$, depicting the piezomagnetic and antipiezomagnetic effects driven by ferro-type and antiferro-type magnetic octupoles in MnF$_2$. For the piezomagnetic effects in (a) and (c), the variations are shown for two different magnetic domains (in green and blue data points) while for anti-piezomagnetic effect the variations of local spin magnetic moments (in green and blue data points) at two Mn atoms are shown. In both cases, variations are also shown for two different strengths of the spin-orbit coupling constant, viz., $\lambda_r=1$ (solid line) and $\lambda_r=2$ (dashed line). The parameter $\lambda_r$ is defined at the caption of Fig. \ref{fig2}.
 }
 \label{fig5}
 \end{figure}

{\it DFT results for MnF$_2$-}
Next, to computationally verify our symmetry-guided prediction of an {\it anti}-piezomagnetic effect and to better understand the microscopic details of both piezo- and antipiezo-magnetic effects, we explicitly study the effect of a shear stress $\sigma_{xz}$ ($\sigma_{yz}$) on the magnetism of MnF$_2$ within the DFT framework. For each value of strain, we relax the internal atomic coordinates while fixing the lattice constants to the strained values. The results of our calculations are depicted in Fig. \ref{fig5}. As is clear from Fig. \ref{fig5} (a) and (c), application of shear stress $\sigma_{xz}$ ($\sigma_{yz}$) generates a net moment along $\hat y$ ($\hat x$), as expected due to the piezomagnetic effect. In addition, as shown in Fig. \ref{fig5} (b) and (d), a tiny spin component appears along $\hat x$ ($\hat y$) at the individual Mn sites with an antiparallel orientation at the neighboring Mn site, corresponding to the predicted {\it anti}-piezomagnetic effect. 

We see from Fig. \ref{fig5} that both piezo- and antipiezo-magnetic responses are linear in nature. Also, in both cases the generated moments reverse their directions in the opposite antiferromagnetic domain. Such a reversal of moment direction is consistent with experimental reports and can be understood from the fact that both ${\cal O}_{32^-}$ and ${\cal O}_{30}$ octupoles have opposite signs in the opposite antiferromagnetic domains. 

Further, to understand the importance of spin-orbit interaction on these effects, we artificially doubled the strength of the spin-orbit coupling in our calculation and as depicted in Fig. \ref{fig5}, this results in an enhancement in the generated moment for both cases. This suggests that, unlike the magnetic Compton scattering described in the next section, both piezo- and antipiezo-magnetic effects are relativistic effects. Physically, this can be understood from the fact that the stress applied to the structure needs to be coupled to the magnetization density of the system, which is mediated via spin-orbit interaction. We note that the dependence on the spin-orbit coupling strength also helps to predict the hierarchy of the piezo and antipiezo-magnetic effects in different materials. For example, the relativistic piezo- and antipiezo-magnetic effects are expected to be much stronger in CoF$_2$ compared to MnF$_2$ due to the strong spin-orbit interaction of the Co atoms in the former.

The predicted anti-piezomagnetic effect should be experimentally observable by detecting the resulting spin canting in the presence of a uniform stress. While the early experiments \cite{Borovikromanov1960}, indeed, indicated rotation of spins upon application of a shear stress so that an antiparallel spin-component is generated in addition to a net magnetization in a piezomagnetic effect, confirmation of the antipiezomagnetic effect would require further measurements to verify the linear generation and the switching of canted moments. 
Another possibility of experimental verification would be to apply a dynamical stress, causing opposite stresses on the two Mn sublattices, so that the  anti-piezomagnetic effect would lead to a net magnetization. Our work, correlating the piezo- and antipiezo-magnetic effects to the magnetic octupoles serves as a guideline for future observation and manipulation of spin arrangements using strain \cite{Disa2020, Formisano2022}.

\subsection{Direct Detection of Magnetic Octupoles: Magnetic Compton Profile in an Antiferromagnet}

The Compton scattering \cite{Compton1923} of x-ray photons, which was an early confirmation of quantum mechanical behavior, is a widely used technique today in fields as diverse as radio-biology, astrophysics, and condensed matter physics. In condensed matter systems, it is used to measure the electron density in momentum space or in an extension known as magnetic Compton scattering, the spin-dependent electron momentum density \cite{PlatzmanTzoar1970}, 
\begin{equation}
 J_{\rm mag} (p_z) = \int \int [\rho_{\uparrow}(\vec p)-\rho_{\downarrow}(\vec p)] dp_x dp_y~~~.  
\end{equation}
Here, $ J_{\rm mag}$ is the magnetic Compton profile (MCP), the key quantity measured in the magnetic Compton scattering measurements, and $\rho_{\uparrow}(\vec p)$ and $\rho_{\downarrow}(\vec p)$ are respectively the up- and down- spin-polarized electron density in momentum space. 

Magnetic Compton scattering has been extensively applied to ferri- and ferro-magnetic systems (with non-zero magnetization) \cite{SakaiOno1976,Cooper1991,Duffy2010,Itou2013,Zukowski1993,Duffy1998,Duffy2000,Banfield2005,ShentonTaylor2007,Duffy2013,Mijnarends2007,Mizoroki2011,Ahuja2013} to
extract spin polarizations at Fermi surfaces \cite{Duffy2013,Mijnarends2007,Mizoroki2011}. In one of our recent works, we proposed that a spin-polarized electron density can also exist in the momentum space of non-magnetic systems, provided that the inversion symmetry is broken, leading to a MCP \cite{Bhowal2022}. To date, however, MCP has not been proposed or measured in conventional antiferromagnets. Because the up and down spin-polarized bands are degenerate, leading to vanishing spin-polarized electron density in momentum space. Here we show that the spin-splitting of the energy bands in antiferromagnets with ferro-type magnetic octupoles results in a non-zero MCP, despite the zero net magnetization. This, in turn, facilitates the MCP as a direct probe for existence of ferro-type magnetic octupoles.

\begin{figure}[t]
\centering
\includegraphics[width=\columnwidth]{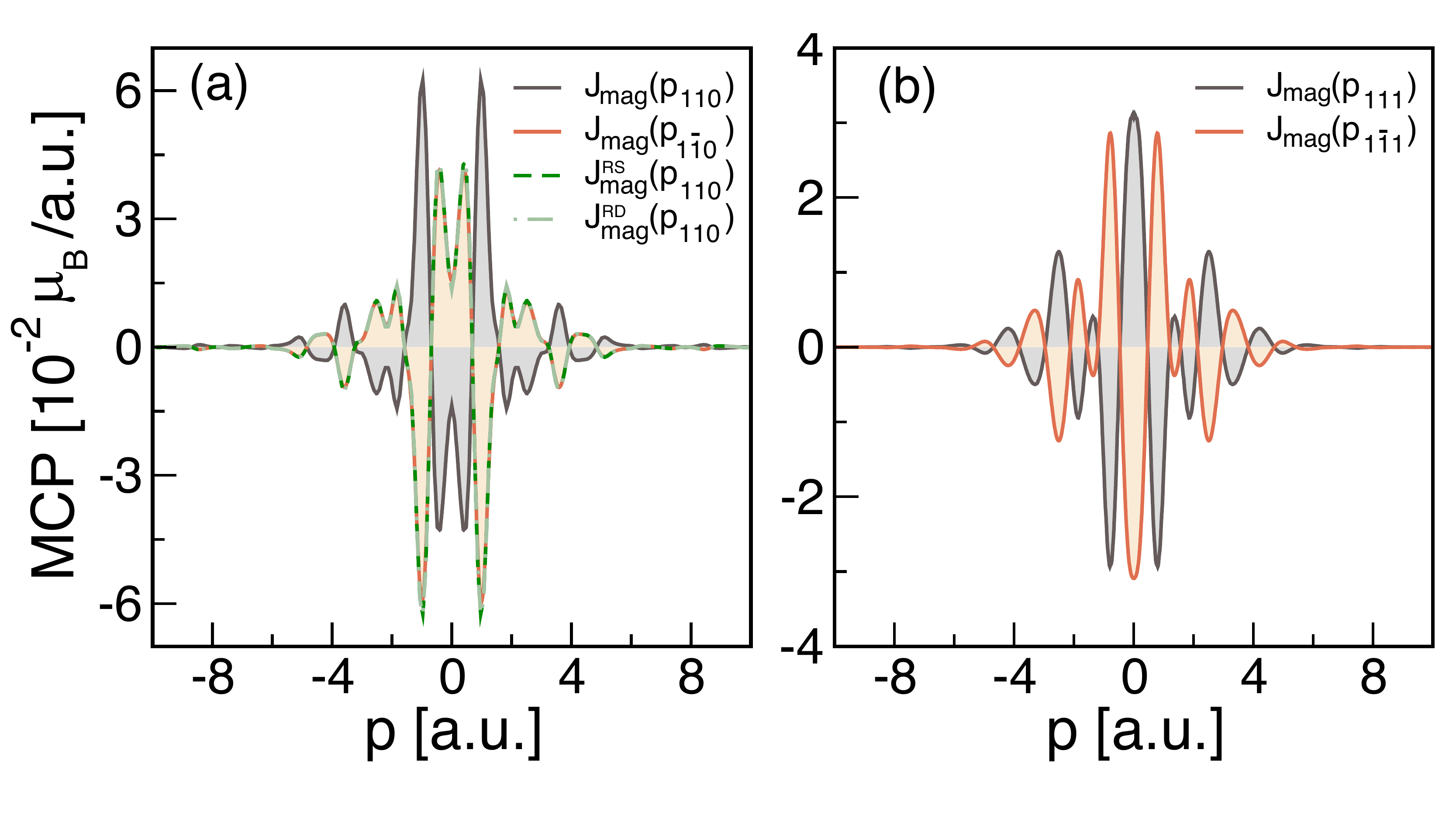}
 \caption{Magnetic Compton profiles (MCPs) of MnF$_2$ along (a) [110] and [1$\bar{1}$0] and (b) [111] and [1$\bar{1}$1] directions in the momentum space. The reversal of the profiles is apparent from (a) and (b) as the momentum direction changes by ${\cal C}_4$ rotation. Panel (a) also depicts the MCPs along [110] direction, $J_{\rm mag}^{\rm RS} (p_{110})$ and $J_{\rm mag}^{\rm RD} (p_{110})$, for the hypothetical modified structure and for the reversed magnetic domain respectively. In both cases, the MCP switches sign compared to the profile of MnF$_2$ along the same momentum direction. 
 }
 \label{fig4}
 \end{figure}

To verify the non-zero MCP for our example material MnF$_2$, we explicitly compute the MCP using the methods implemented in the ELK code (see the computational details in Appendix A). The computed MCP of MnF$_2$ along the [110] direction in momentum space is shown in Fig. \ref{fig4}(a). This is to our knowledge the first identification of a MCP in an AFM. We note that the MCP is present even without including spin-orbit effects, as expected due to the non-relativistic spin splitting in MnF$_2$. Note also that the integral of the MCP is zero, consistent with the net vanishing moment in the system.

The characteristics of the computed MCP are quite different from those of nonmagnetic ferroelectrics. First, the computed MCP is symmetric in $\vec p$ in contrast to the antisymmetric MCP in ferroelectrics \cite{Bhowal2022}. This follows from the symmetric and antisymmetric spin splitting in MnF$_2$ and ferroelectrics respectively. More importantly, however, the magnitude of the MCP in MnF$_2$ is larger by about an order of magnitude compared to the computed values for the ferroelectrics, PbTiO$_3$ and GeTe \cite{Bhowal2022}. This again is associated with the large NRSS of the bands in contrast to the weak relativistic spin-splitting of the bands in ferroelectrics. Finally, as shown in Fig. \ref{fig4} (a), the MCP in MnF$_2$ changes sign as the momentum direction is changed from the [110] to the [1$\bar{1}$0] direction, unlike the case of ferroelectrics for which the profile, being antisymmetric, switches sign as $\vec p \rightarrow -\vec p$. Such sign reversal of the MCP in MnF$_2$ is understandable from the reversal of the spin splitting as the momentum direction changes from [110] to [1$\bar{1}$0] (see Fig. \ref{fig3} (b)).

Since the magnetic octupole leads to the spin-splitting of the bands, which, in turn, gives rise to the MCP, the MCP provides a direct measurement of the existence of ferroically ordered magnetic octupoles in MnF$_2$. For further confirmation we compute the MCPs for the cases of the reversed structure and the other AFM domain (described in section \ref{manipulation}), for both of which the ferro-type ${\cal O}_{32^-}$ magnetic octupole reverses sign. Indeed, the computed MCPs, shown in Fig. \ref{fig4} (a), reverse the sign of their profile, in agreement with our expectation.  

{\it Proposed Experimental Setup.} The measurement setup needed to detect magnetic octupoles using MCP will be similar to that of a conventional magnetic Compton scattering experiment with circularly polarized light. Generally, the measurements are performed in back-scattering geometry with either parallel spin and momentum directions or along a momentum direction that has at least one component along the direction of the spin polarization. Since the spin polarization direction in MnF$_2$ is along $\hat z$, we further compute the MCPs along the [111] direction in reciprocal space. As depicted in Fig. \ref{fig4} (b), the computed MCP, although smaller compared to that along [110], still has a much larger magnitude compared to the case of a ferroelectric, suggesting that it is likely discernible in experiments. 

We note that since the two antiferromagnetic domains lead to opposite spin splitting, it is crucial to carry out the measurements on a single antiferromagnetic domain of MnF$_2$. Such a single antiferromagnetic domain can be obtained by the simultaneous application of a uniaxial stress and a magnetic field while cooling the sample through the N{\'e}el temperature T$_{\rm N} \approx$ 67 K \cite{Baruchel1980,Baruchel1988}.  It is interesting to point out here that the combination of stress and magnetic field is in fact the conjugate field of a magnetic octupole, and as described before each of the antiferromagnetic domains can indeed be identified as a ferro-octupolar domain. Such a single magnetic domain is also referred to as a piezomagnetic domain due to its close connection to the piezomagnetic effect in MnF$_2$ \cite{Baruchel1980}, driven by the ferroic magnetic octupoles as discussed in the previous section.

\section{Discussions and Outlook} \label{discussion}

To summarize, using MnF$_2$ as an example material, we have shown that the order parameter of centrosymmetric antiferromagnets with NRSS is the magnetic octupole, since it is the lowest-order ferroically ordered magnetic quantity in this case. The ferromagneto-octupolar ordering provides a convenient description of the NRSS and reveals the conjugate field -- a combined shear stress and magnetic field -- which can in turn manipulate the NRSS through selection of the magnetic domain. The magnetic octupole description explains the reported piezomagnetic response of such systems, and allows us to predict an as-yet-unobserved nonlinear magnetoelectric effect as well as an antipiezomagnetic effect resulting from an additional antiferroic arrangement of different magnetic octupoles. Finally, we propose magnetic Compton scattering for the direct detection of magnetic octupoles in such unconventional antiferromagnets. 

We note that centrosymmetric antiferromagnets with NRSS may also have higher-order ferroically ordered even-parity magnetic multipoles in addition to their ferromagneto-octupolar order. These higher-order multipoles are relevant for describing NRSS with $g$-wave or $i$-wave symmetry. For example, Fe$_2$O$_3$ in its low-temperature state with magnetic moments oriented along the symmetry axis, which is reported to have a $g$-wave spin splitting \cite{Smejkal2022}, allows for a magnetic triakontadipole in addition to the magnetic octupole. The connection between this rank-5 even-parity magnetic multipole and the corresponding $g$-wave NRSS is an interesting direction for future study.  

In addition to providing important insight into the newly discovered unconventional antiferromagnets with NRSS, the results presented here are relevant for the prolonged effort to reveal and detect the magnetic octupolar phase \cite{Santini2000, Kuramoto2000, Kusunose2007, Matsumura2009, Suzuki2017, Higo2018, Patri2019, Maharaj2020, Khaliullin2021, Kimata2021, Urru2022,Sreekar2022}, as well as for potential applications through strain engineering of antiferromagnetism via the piezo- or antipiezo-magnetic effect in spintronic devices.  We note that magnetic octupoles are also likely to be relevant for the reported spin-phonon interaction \cite{Lockwood1988,Cottam2019} and surface magnetization \cite{Nizhankovskii2000} in MnF$_2$, and could shed light on the reported strong magnetic anisotropy in doped FeSb$_2$ \cite{Mazin2021}. 

Merging the seemingly disconnected fields of hidden order, antiferromagnetic spintronics, and inelastic scattering techniques, our work opens up new directions for exploration which we hope will motivate both theoretical and experimental investigation in the near future.

\section*{Acknowledgements}
We thank Steve Collins, Jon Duffy, Urs Staub, and Andrea Urru for stimulating discussions.  
NAS and SB were supported by the ERC under the EU’s Horizon 2020 Research and Innovation Programme grant No 810451 and by the ETH Zurich. Computational resources were provided by ETH Zurich's Euler cluster, and the Swiss National Supercomputing Centre, project ID eth3. 

\section*{Appendix}
\subsection{Computational Details}

The electronic structure of MnF$_2$ has been computed using the linearized augmented plane wave (LAPW) method as implemented in the ELK code \cite{code,elk}. We use the LDA+SOC+$U$ formalism, with $U_{\rm eff}=5$ eV at the Mn site \cite{Yuan2020}. A basis set of $l_{max(apw)} = 8$, a $5\times5\times7$ k-point sampling of the Brillouin zone are used to achieve self-consistency. The product of the muffin-tin radius (2.4, and 2 a.u. for Mn and F respectively) and the maximum reciprocal lattice vector is taken to be 7. The magnetic Compton profile and the atomic-site multipoles are computed using the extended versions \cite{Ernsting2014,Spaldin2013} of the Elk code. The spin-polarized electron momentum densities are calculated and projected onto the selected momentum directions ($\vec p$) to obtain the magnetic Compton profile following the implementations, reported in Ref. \cite{Ernsting2014}. The computed MCP is scaled to the factor that normalizes the valence contribution of the total Compton profile to the total number of valence electrons per formula unit of MnF$_2$ in the calculation, which is 29 in this case. For the computation of atomic-site multipoles, the density matrix $\rho_{lm,l'm'}$ is decomposed into the tensor moments, of which the parity even tensor moments have contributions from $l=l'$ terms. We, therefore, evaluate both $d-d$ and $p-p$ matrix element contributions for the multipoles at the Mn site. 
The electronic structure of MnF$_2$ is also computed within the plane-wave based projector augmented wave (PAW) \cite{Bloch1994,Kresse1999} method as implemented in the Vienna ab initio simulation package (VASP) \cite{Kresse1993,Kresse1996} and the results agree well with that computed using the ELK code. The atomic relaxations in presence of shear strain in the piezomagnetic effect are carried out until the Hellman-Feynman forces on each atom becomes less than 0.01 eV/\AA. 

\begin{figure}[t]
\centering
\includegraphics[width=\columnwidth]{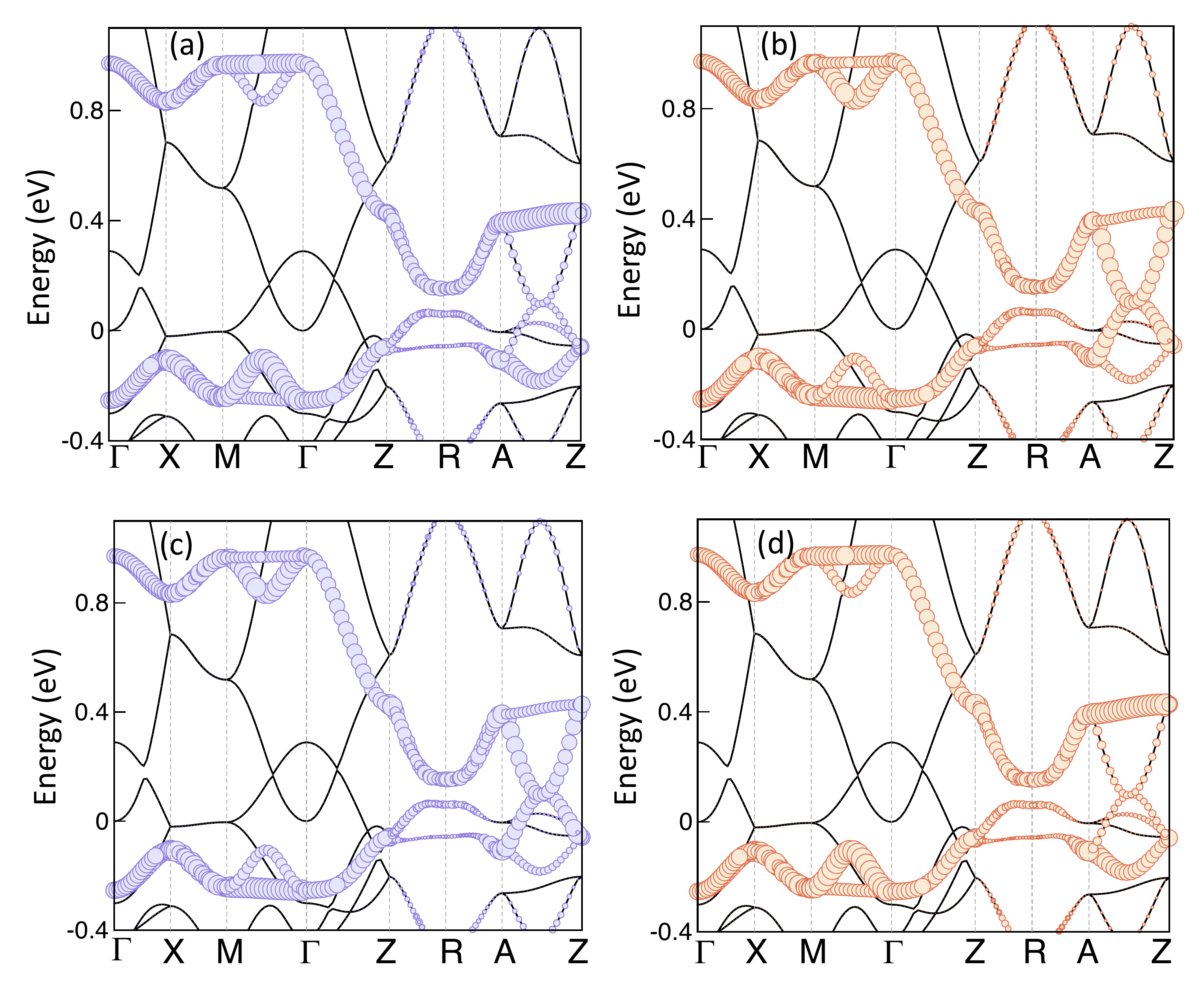}
 \caption{Nonmagnetic band structure of MnF$_2$, showing the $d_{xz}$ and $d_{yz}$ orbital contributions for the (a) Mn1 and (b) Mn2 sublattices. (c) and (d) depict the same for the hypothetical modified structure, indicating a reversal of Mn1 and Mn2 sublattice contributions for the modified structure.
 }
 \label{fig6}
 \end{figure}

\subsection{Tight-binding Parameters}

The realistic tight-binding parameters of the Hamiltonian (\ref{TB}), i.e., the effective $d-d$ hoppings $t_i$ ($i=1,4$) and the onsite energies $\varepsilon_i$ ($i=1,2$) in Eq. \ref{functions} are extracted from the DFT calculations by downfolding the effect of the F-$p$ orbitals using the N$^{th}$ order muffin-tin orbital (NMTO) method \cite{AndersenSaha-Dasgupta}. The computed parameters are listed in Table \ref{tab1}.
\begin{table} [t]
\caption{Tight-binding parameters of the Hamiltonian (\ref{TB}), extracted using NMTO method.}
\setlength{\tabcolsep}{4pt}
\centering
\begin{tabular}{ c| c | c |   c| c| c}
\hline
\multicolumn{4}{c}{$d$-$d$ hopping parameters (Ry)}  \ & \multicolumn{2}{|c}{Onsite energies (Ry)}  \\
$t_1$ \ & $t_2$  \ & $t_3$ \ & $t_4$   \ &   $\varepsilon_1$ & $\varepsilon_2$ \\ [1 ex]
\hline
0.0036 & -0.0038  & 0.0040 & 0.0034 & -0.1385 & -0.0339 \\
  \hline
\end{tabular}
\label{tab1}
\end{table}

\subsection{DFT band structures in absence of magnetism}

The computed band structure in absence of magnetism is shown in Fig. \ref{fig6}, depicting the splitting between bands of two different sublattice contributions along $\Gamma \rightarrow$ M. For example, for the pair of bands around 1 eV along $\Gamma \rightarrow$ M, the top band has predominant contributions from the Mn1 sublattice while the bottom band is predominantly of Mn2 sublattice character. For the modified structure, described in Section \ref{manipulation}, the computed atom and orbital projected band structure shows that the band structure remains identical except that the sublattice characters of the same pair of bands are reversed.

\bibliographystyle{apsrev4-1}
\bibliography{LNPO}

\end{document}